\documentclass{aa}
\usepackage{graphicx}
\usepackage{txfonts}
\begin{document}
   \title{The correlation between C/O ratio, metallicity and the
initial WD mass for SNe Ia}

%   \subtitle{I. Overviewing the $\kappa$-mechanism}

   \author{Xiangcun Meng
          \inst{1}
          and
          Wuming Yang \inst{1,2}}
%\fnmsep\thanks{Just to show the usage
%          of the elements in the author field}

   \offprints{X. Meng}

   \institute{School of Physics and Chemistry, Henan Polytechnic
University, Jiaozuo, 454000, China\\
              \email{xiangcunmeng@hotmail.com}
           \and
Department of Astronomy, Beijing Normal University, Beijing
100875, China
%             \email{c.ptolemy@hipparch.uheaven.space}
%             \thanks{The university of heaven temporarily does not
%                     accept e-mails}
             }
   \date{Received; accepted}

% \abstract{}{}{}{}{}
% 5 {} token are mandatory

  \abstract
   {When type Ia supernovae (SNe Ia) were chosen as distance indicator
to measure cosmological parameters, Phillips relation was applied.
However, the origin of the scatter of the maximum luminosity of
SNe Ia (or the variation of the production of $^{\rm 56}$Ni) is
still unclear. The metallicity and the carbon abundance of white
dwarf (WD) before supernova explosion are possible key, but
neither of them has an ability to interpret the scatter of the
maximum luminosity of SNe Ia.}
   {In this paper, we want to check whether or not the carbon
abundance can be affected by initial metallicity.}
   {We calculated a series of stellar evolution.}
   {We found that when $Z\leq0.02$, the carbon abundance is
   almost independent of metallicity if it is plotted against the initial WD
mass. However, when $Z>0.02$, the carbon abundance is not only a
function of the initial WD mass, but also metallicity, i.e. for a
given initial WD mass, the higher the metallicity, the lower the
carbon abundance. Based on some previous studies, i.e. both a high
metallicity and a low carbon abundance lead to a lower production
of $^{\rm 56}$Ni formed during SN Ia explosion, the effects of the
carbon abundance and the metallicity on the amount of $^{\rm
56}$Ni are enhanced by each other, which may account for the
variation of maximum luminosity of SNe Ia, at least
qualitatively.}
   {Considering that the central density of WD before
supernova explosion may also play a role on the production of
$^{\rm 56}$Ni and the carbon abundance, the metallicity and the
central density are all determined by the initial parameters of
progenitor system, i.e. the initial WD mass, metallicity, orbital
period and secondary mass, the amount of $^{\rm 56}$Ni might be a
function of the initial parameters. Then, our results might
construct a bridge linking the progenitor model and the explosion
model of SNe Ia.}

   \authorrunning{Meng \& Yang}
   \titlerunning{The correlation between C/O ratio, metallicity and WD mass}
   \maketitle{}
%
%________________________________________________________________

\section{Introduction}\label{sect:1}
In their function as one of the distance indicators, type Ia
supernovae (SNe Ia) show their importance in determining
cosmological parameters, which resulted in the discovery of the
accelerating expansion of the universe (Riess et al. \cite{REI98};
Perlmutter et al. \cite{PER99}). The result was exciting and
suggested the presence of dark energy. At present, SNe Ia are
proposed to be cosmological probes for testing the evolution of
the dark energy equation of state with time and testing the
evolutionary history of the universe (Riess et al. \cite{RIESS07};
Kuznetsova et al. \cite{KUZNETSOVA08}; Howell et al.
\cite{HOWEL09}).

When SNe Ia are applied as a distance indicator, the Phillips
relation is adopted, which is a linear relation between the
absolute magnitude of SNe Ia at maximum light and the magnitude
drop of the B light curve during the first 15 days following the
maximum (Phillips \cite{PHI93}). This relation was motivated by
the observations of two peculiar events, i.e. SN 1991bg and SN
1991T, and implies that the brightness of SNe Ia is mainly
determined by one parameter. There is a consensus that a SN Ia is
from a thermonuclear explosion of a carbon-oxygen white dwarf (CO
WD) and the amount of $^{\rm 56}$Ni formed during the supernova
explosion dominates the maximum luminosity of SNe Ia (Arnett
\cite{ARN82}). But the origin of the variation of the amount of
$^{\rm 56}$Ni for different SNe Ia is still unclear (Podsiadlowski
et al. \cite{POD08}). Many efforts have been paid to resolve this
problem. Some multi-dimensional numerical simulations showed that
the ignition intensity (the number of ignition points) in the
center of WDs or the transition density from deflagration to
detonation are wonderful parameters interpreting the Phillips
relation (Hillebrandt \& Niemeyer \cite{HN00}; H\"{o}flich et al.
\cite{HOFLIC06, HOFLIC10}; Kasen et al. \cite{KASEN10}). In
addition, the ratio of nuclear-statistical-equilibrium (NES) to
intermediate-mass elements (IME) in the explosion ejecta is likely
the key parameter for the width of SN Ia light curve and its peak
luminosity (Pinto \& Eastman \cite{PINTO01}; Mazzali et al.
\cite{MAZZALI01,MAZZALI07}). For the simulations above, a
reasonable assumption is that these parameters are determined by
one or some properties of SNe Ia progenitor. Since these
parameters are free ones for the numerical simulations of SNe Ia
explosion, the question was turned into which property or
properties of progenitor system determine these parameters. This
is still unclear. Lesaffre et al. (\cite{LESAFFRE06}) carried out
a systematic study of the sensitivity of ignition conditions for
H-rich Chandra single degenerate exploders on various properties
of the progenitors, and suggested that the central density of the
WD at ignition may be the origin of the Phillips relation (see
also Podsiadlowski et al. \cite{POD08}). This suggestion was
uphold by detailed multi-dimensional numerical simulations of
explosion (Krueger et al. \cite{KRUEGER10}). But in the models of
H\"{o}flich et al. (\cite{HOFLIC10}), the central density is only
a second parameter, and the cooling time of the WDs before mass
transfer in Lesaffre et al. (\cite{LESAFFRE06}) and Krueger et al.
(\cite{KRUEGER10}) is shorter than 1 Gyr. However, there are SNe
Ia as old as 10 Gyr. The WDs with such a long cooling time may
become more degenerate before the onset of accretion phase. Some
other processes like C and O separation or crystallization may
occur, and dominate the properties of the CO WD (Fontaine et al.
\cite{FONTAINE01}). How the extremely degenerate conditions affect
the properties of SNe Ia is still unclear. Then, the suggestion of
Lesaffre et al. (\cite{LESAFFRE06}) should be checked carefully
under extremely degenerate conditions (Bravo et al.
\cite{BRAVO10b}). Some numerical and synthetical studies showed
that metallicity has an effect on the final amount of $^{\rm
56}$Ni, and thus the maximum luminosity (Timmes et al.
\cite{TIM03}; Travaglio et al. \cite{TRA05}; Podsiadlowski et al.
\cite{POD06}; Bravo et al. \cite{BRAVO10}) and there do be some
observational evidence of the correlation between the properties
of SNe Ia and metallicity (Branch \& Bergh \cite{BB93}; Hamuy et
al. \cite{HAM96}; Wang et al. \cite{WAN97}; Cappellaro et al.
\cite{CAP97}; Shanks et al. \cite{SHA02}). However, the
metallicity seems not to have the ability to interpret the whole
scatter of the maximum luminosity of SNe Ia (Timmes et al.
\cite{TIM03}; Gallagher et al. \cite{GALLAGHER08}; Howell et al.
\cite{HOWEL09b}).

Nomoto et al. (\cite{NOM99, NOM03}) suggested that the
average\footnote{If there is no special statement, the C/O ratio
in this paper is the average value over the whole WD structure
just before the SN explosion.} ratio of carbon to oxygen (C/O) of
a white dwarf at the moment of explosion is the dominant parameter
for the Phillips relation (see also Umeda et al. \cite{UMEDA99}).
The higher the C/O, the larger the amount of nickel-56, and then
the higher the maximum luminosity of SNe Ia. By comparing theory
and observations, the results of Meng, Chen \& Han
(\cite{MENGXC09}) and Meng \& Yang (\cite{MENGYANG10a}) upheld
this suggestion. Nomoto et al. (\cite{NOM99, NOM03}) used total
$^{\rm 12}$C mass fraction included in the convective core of mass
$M=1.14 M_{\odot}$ just before the SN Ia explosion, X(c), to
represent the C/O ratio and their suggestion is based on a fact
that the dependence of X(c) on metallicity is small when it is
plotted against the initial mass of WDs when $Z\leq0.02$.
Actually, when $Z\leq0.02$, the effect of metallicity on the
amount of $^{\rm 56}$Ni can be neglected. Only when $Z>0.02$,
metallicity has a significant influence on the production of
$^{\rm 56}$Ni in SNe Ia explosion (Timmes et al. \cite{TIM03}).
Here, we want to check whether or not the C/O ratio still can not
be affected by metallicity when $Z>0.02$.

In section \ref{sect:2}, we simply describe our method, and
present the calculation results in section \ref{sect:3}. In
section \ref{sect:4}, we show discussions and our main
conclusions.

%__________________________________________________________________

\section{METHOD}\label{sect:2}

%                                     Two column figure (place early!)
%______________________________________________ Gamma_1 (lg rho, lg e)
In this paper, our work is based on the single degenerate
scenario, i.e. a SN Ia is from a CO WD in a binary system and its
companion is a main-sequence or a slightly evolved star (WD+MS), a
red giant star (WD+RG) or a helium  star (WD + He star) (Whelan \&
Iben \cite{WI73}; Nomoto, Thielemann \& Yokoi \cite{NTY84}). This
scenario has been widely studied by many groups (Li \& van den
Heuvel \cite{LI97}; Hachisu et al. \cite{HAC99a}; Langer et al.
\cite{LAN00}; Han \& Podsiadlowski \cite{HAN04}; Chen \& Li
\cite{CHENWC07}; Hachisu et al. \cite{HKN08}; Meng, Chen \& Han
\cite{MENGXC09}; L\"{u} et al. \cite{LGL09}; Wang et al.
\cite{WANGB09a,WANGB09b}; Meng \& Yang
\cite{MENGYANG10b,MENGYANG10c}; Wang, Li \& Han \cite{WANGB10}).
We assume that an initial CO WD is derived from a main sequence
(MS) star in a primordial binary system (primary). The CO WD
accretes hydrogen-rich material from its companion via Roche lobe
overflow or wind, where the companion is a normal star. The
accreted hydrogen-rich material is burned into helium, and then
the helium is converted to carbon and oxygen. The CO WD increases
its mass until the mass reaches 1.378 $M_{\odot}$(close to the
Chandrasekhar mass limit, Nomoto, Thielemann \& Yokoi
\cite{NTY84}) where it explodes in a thermonuclear supernova. A
binary systems with the same primordial primary but different
orbital period may produce a CO WD with different mass. For
simplicity, we assume that if the MS mass of the primary in a
primordial binary system is same, the initial mass of the CO WD
from the binary system is same and is equal to the core mass
derived from the envelope-ejection model in Han, Podsiadlowski \&
Eggleton (\cite{HAN94}) and Meng et al. (\cite{MENG08}) (see below
in details). Because the primary in a binary system may lose its
hydrogen envelope due to the influence of secondary before it
fulfills the criterion for the envelope ejection, the initial mass
of CO WD in this paper should be taken as an upper limit for a
real case. Actually, this assumption can not affect our results
since we only want to find a relation between the C/O ratio, the
WD mass and metallicity. The method used here is similar to that
in Umeda et al. (\cite{UME99}) and H\"{o}flich et al.
(\cite{HOFLIC10}).

The C/O ratio before SN Ia explosion is a result of stellar and
binary evolution, i.e. during central helium burning and thin
shell burning during the stellar evolution and the accretion to
close to the Chandrasekhar mass limit (1.378 $M_{\odot}$, Nomoto,
Thielemann \& Yokoi \cite{NTY84}). After the central helium
burning phase, the C/O ratio is low, i.e. $0.25-0.5$ depending on
the initial mass and metallicity of main sequence (MS) star (Umeda
et al. \cite{UME99}). The C/O ratio obtained from the burning
shell is $\approx1$, because the helium in the shell has a lower
density and a higher temperature compared to helium burning in the
core (H\"{o}flich et al. \cite{HOFLIC10}).

In the paper, we calculate a series of stellar evolutions with the
primordial MS mass from 1 $M_{\odot}$ to 6.5 $M_{\odot}$ until the
stars evolve to the asymptotic giant branch (AGB) stage. When a
star evolves to the stage, its envelope may be lost if the binding
energy (BE) of the envelope transforms from a negative phase to
positive one (Paczy\'{n}ski \& Zi\'{o}lkowski \cite{PZ68}, see
also Fig. 2 in Meng et al. \cite{MENG08}). We calculated the BE of
the envelope by
  \begin{equation}
 \Delta W=\int_{M_{\rm c}}^{M_{\rm s}}(-\frac{Gm}{r}+U){\rm d}m,
 \label{eqbe}
  \end{equation}
where $M_{\rm c}$ is the core mass, $M_{\rm s}$ is the surface
value of the mass coordinate $m$. $U$ is the internal energy of
thermodynamics where those due to ionization of H and dissociation
of $\rm H_{\rm 2}$, as well as the basic $\frac{3}{2}\Re T/\mu$
for a perfect gas are all included). Here, we assume that a star
will lose its envelope when the BE of the star's envelope
increases to the point of $\Delta W=0$ and the core mass at the
point is the final WD mass. The method here is robust and its
virtue is significant because we need not consider the specific
mechanism of mass loss since the mass loss rate is very
uncertainty (see Meng et al. \cite{MENG08} in details about this
method). We assume that the remnant after envelope ejection is a
CO WD if carbon and oxygen have not been ignited at the moment of
envelope ejection. Following shell burning, we assume that the C/O
ratio is 1\footnote{In fact, for the core chemical profile
produced during the AGB phase, the steady increase of the carbon
abundance in the region adjacent to the inner flat profile is
synthesized by the shell in the early AGB, whereas the following
abrupt peak is left by the thermally pulsing phase. The slop of
the profile produced during the early AGB depends on both the
initial mass and metallicity, and the C/O$=1$ is not always a good
approximation. However, to compare with previous results, we still
simply make a constant-ratio assumption.} until $M_{\rm
WD}=1.378M_{\odot}$ as did in Umeda et al. (\cite{UME99}) and
H\"{o}flich et al. (\cite{HOFLIC10}).

We use the stellar evolution code of Eggleton
(\cite{EGG71,EGG72,EGG73}), which has been updated with the latest
input physics over the last three decades (Han, Podsiadlowski \&
Eggleton \cite{HAN94}; Pols et al. \cite{POL95,POL98}). The
chemistry of a WD is mainly determined by the competition of two
major nuclear reactions powering the He burning, i.e. $3\alpha$
and $^{\rm 12}{\rm C}(\alpha,\gamma)^{\rm 16}{\rm O}$. As
discussed by Imbriani et al. (\cite{IMBRIANI01}) and Prada Moroni
\& Straniero (\cite{PRADA02}), the final C/O after the central
helium exhaustion not only depends on the rate of these two
reactions, but also is significantly influenced by the efficiency
of convective mixing operating at the central helium burning
phase. We set the ratio of mixing length to local pressure scale
height, $\alpha=l/H_{\rm p}$, to 2.0, and set the convective
overshooting parameter, $\delta_{\rm OV}$, to 0.12 (Pols et al.
\cite{POL97}; Schr\"{o}der et al. \cite{SCH97}), which roughly
corresponds to an overshooting length of $0.25 H_{\rm P}$. The two
parameters are adopted during the whole evolution of star. For the
occurrence of convective instability near the He exhaustion in the
central core, some breathing pulses are expected. (Castellani
\cite{CASTELLANI85, CASTELLANI89}). This phenomenon may also occur
naturally in our code. The reaction rates are from Caughlan \&
Fowler (\cite{CAUGHLAN88}), except for the $^{\rm 12}{\rm
C}(\alpha,\gamma)^{\rm 16}{\rm O}$ reaction which is taken from
Caughlan et al. (\cite{CAUGHLAN85}). The range of metallicity is
from 0.0001 to 0.1, i.e. 0.0001, 0.0003, 0.001, 0.004, 0.01, 0.02,
0.03, 0.04, 0.05, 0.06, 0.08, 0.1. The opacity tables for these
metallicties are compiled by Chen \& Tout (\cite{CHE07}) from
Iglesias \& Rogers (\cite{IR96}) and Alexander \& Ferguson
(\cite{AF94}). For a given $Z$, the initial hydrogen mass fraction
is assumed by
 \begin{equation}
 X=0.76-3.0Z,
\label{eqhy}
  \end{equation}
(Pols et al. \cite{POL98}), and then the helium mass fraction is
$Y=1-X-Z=0.24+2Z$. Based on the correlation between $X$, $Y$ and
$Z$ used here, Pols et al. (\cite{POL98}) accurately reproduced
the color-magnitude diagrams (CMD) of some clusters.

In this paper, we skip thermal pulses by taking a longer time-step
to reduce computing time, i.e we study the average evolution of
thermally pulsing AGB models. This treatment on the thermal pulses
may lose some information about the structural change of the
envelope due to thermal pulse. However, these treatment will not
affect our final results because the situation of $\Delta W=0$ is
fulfilled before the onset of thermal pulse and after the point a
simple assumption of $C/O=1$ is adopted (see also in Meng et al.
\cite{MENG08}).

   \begin{figure}
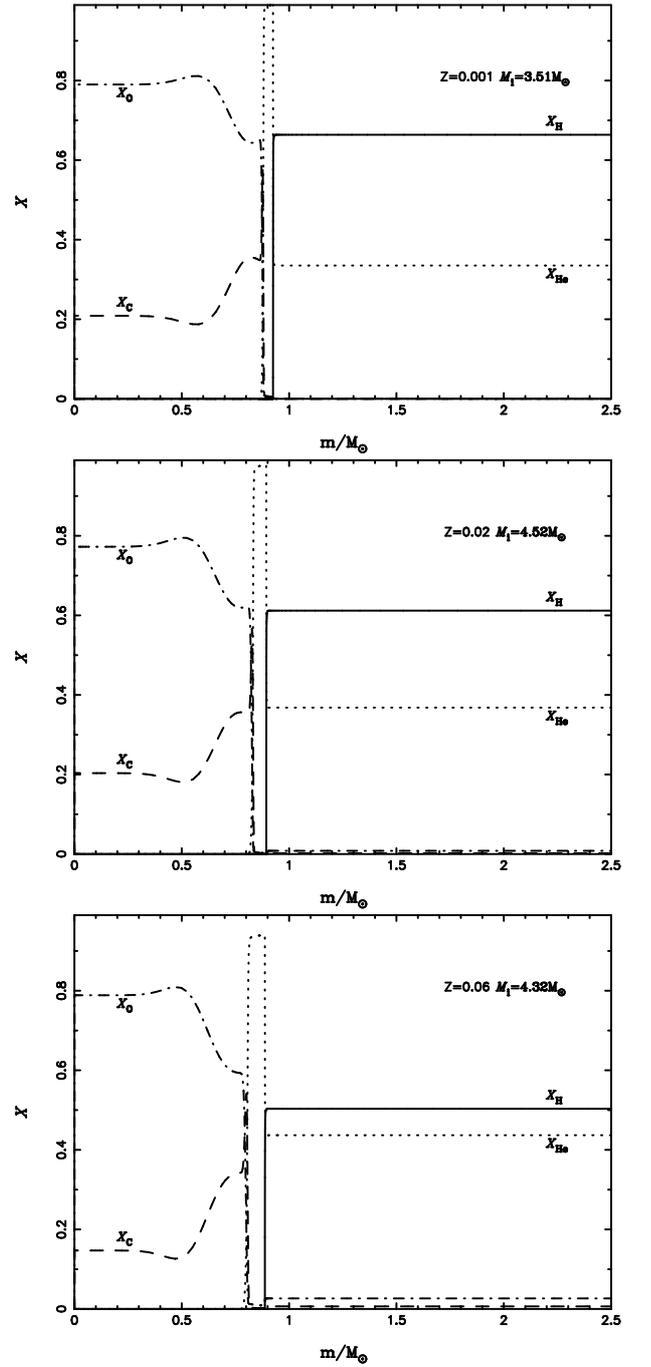

   \centering
   \includegraphics[width=60mm,height=80mm,angle=270.0]{001.ps}
   \includegraphics[width=60mm,height=80mm,angle=270.0]{02.ps}
   \includegraphics[width=60mm,height=80mm,angle=270.0]{06.ps}
   \caption{Abundances of several elements in mass fraction in the
inner core of three models with different primordial MS masses and
different metallicities at the moment of envelope ejection. The
solid, dotted, dashed and dot-dashed lines represent hydrogen,
helium, carbon and oxygen abundance, respectively. Top: $Z=0.001$
and $M_{\rm i}=3.51$; Middle: $Z=0.02$ and $M_{\rm i}=4.52$;
Bottom: $Z=0.06$ and $M_{\rm i}=4.32$.}
              \label{structure}%
    \end{figure}

\section{Results}\label{sect:3}
In this paper, we use total carbon abundance  to represent the C/O
ratio and a core mass to represent the initial WD mass as did in
Umeda et al. (\cite{UMEDA99}). In figure \ref{structure}, we show
the abundances of several elements in mass fraction in the inner
core of three representative models with different metallicity at
the moment of envelope ejection. In the figure, the H burning
shell is located around the mass coordinate of $m=0.9 M_{\odot}$,
which means that the remnants from the models have a similar mass,
i.e. the initial WD mass from the stars is similar. An interesting
feature in the figure is that the carbon abundance and the flat
part of the carbon abundance profile in the inner core decrease
with metallicities (Umeda et al. \cite{UME99}; Dominguez et al.
\cite{DOMINGUEZ01}).The carbon abundance is mainly derived from
the result of the competition between $3\alpha$ and $^{\rm 12}{\rm
C}(\alpha,\gamma)^{\rm 16}{\rm O}$ reaction in the central helium
burning phase, while the flat profile is the result of the central
helium burning, which occurs in a convective core. An increase of
the metallicity leads to an increase of the radiative opacity, and
consequently, a decrease of the central temperature at given
phase, the decrease of the He core mass at the $3\alpha$ onset.
For the central helium burning phase, this leads to a smaller
convective core, hence a smaller region characterized by a flat
C/O profile. A low central temperature favors the destruction
reaction of $^{\rm 12}$C, namely the $^{\rm 12}{\rm
C}(\alpha,\gamma)^{\rm 16}{\rm O}$ with respect the $3\alpha$
reaction, which is the main cause for the decrease of the central
carbon abundance. In addition, the central carbon abundance is
also relevant to the helium abundance in central helium burning
phase, and then to metallicity via $Y=1-Z$ at the beginning of
central helium burning. For a given temperature, a low helium
abundance, i.e. a high metallicity, means a slight lower burning
rates of $3\alpha$ reaction. However, between the model of
$Z=0.001$ and $Z=0.02$ in figure \ref{structure}, the difference
of carbon abundance is not significant, but the difference between
the model of $Z=0.02$ and $Z=0.06$ is remarkable. Then, the effect
of metallicity on the C/O ratio may not be neglected when
$Z>0.02$. In addition, the hydrogen abundance decreases and helium
abundance increase with metallicity in the figure, which is a
natural result of equation \ref{eqhy}\footnote{The hydrogen
abundance and the helium abundance are not accurately equal to
those derived from equation \ref{eqhy} because of the first and
second dredge-up (Busso et al. \cite{BUSSO99}).}.

%
%
% may be higher than that in Meng \& Yang (\cite{MENGYANG10a}) since some special effects
% such as mass-stripping effect and the effect of thermal unstable disk are considered
% in Meng \& Yang (\cite{MENGYANG10a}).}\footnote{Here, the gray points are plotted based on the study in
%Meng et al}

   \begin{figure}
   \centering
   \includegraphics[width=60mm,height=80mm,angle=270.0]{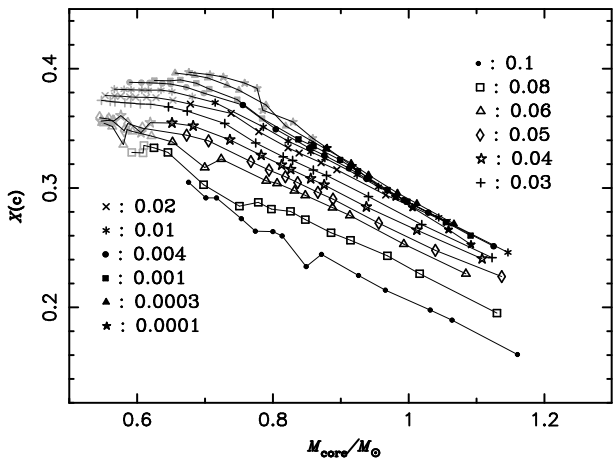}
   \caption{The relation between the total carbon abundance
   of WDs before supernova explosion and the initial
WD mass for SNe Ia with different metallicities. The gray points
represent those that may not contribute to SNe Ia.}
\label{xc}%
    \end{figure}

       \begin{figure}
   \centering
   \includegraphics[width=60mm,height=80mm,angle=270.0]{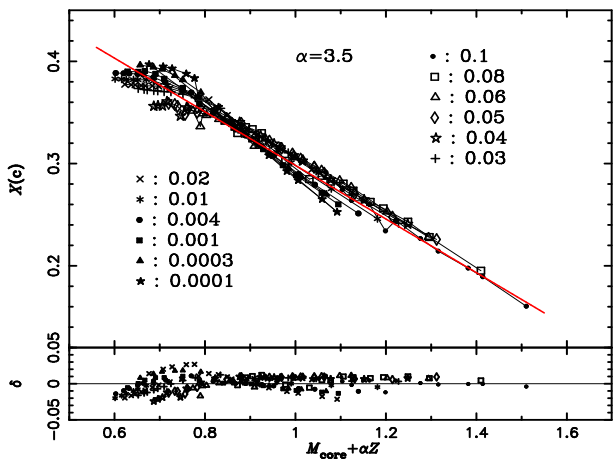}
   \caption{The relation between the total carbon abundance
   of WDs before supernova explosion, the initial
WD mass and metallicity. The solid line is the best fitted linear
relation, where the line fits all the points shown in Fig.
\ref{xc}. The lower panel shows the difference between the points
and the fitted line.} \label{fundz350}%
    \end{figure}

In figure \ref{xc}\footnote{Here, the gray points are plotted
based on the study in Meng et al (\cite{MENGXC09}). Please keep in
mind that the results for the lower limit of WDs for SNe Ia in
Meng et al. (\cite{MENGXC09}) may be higher than that in Meng \&
Yang (\cite{MENGYANG10a}) since some special effects such as
mass-stripping effect of an optically thick wind and the effect of
thermal unstable disk are considered in Meng \& Yang
(\cite{MENGYANG10a}).}, we show the relation between the total
carbon mass fraction of WDs before SNe Ia explosion and the
initial WD mass for SNe Ia with different metallicities. For the
cases of $Z\leq0.02$, the results here is similar to that in
Nomoto et al. (\cite{NOM99, NOM03}), i.e. the relation between the
carbon abundance and the initial WD mass is independent of
metallicity, especially for $M_{\rm WD}\geq0.8 M_{\odot}$. When
$M_{\rm WD}\leq0.8 M_{\odot}$, the scatter is enlarged for the
cases of $Z\leq0.02$, which is mainly from low-metallicity cases
($\leq0.001$, see also Fig. 12 in Umeda et al. \cite{UME99}).
However, for the low-metallicity cases, the WDs with a mass
smaller than $0.8 M_{\odot}$ may not contribute to SNe Ia (see
figure 5 in Meng, Chen \& Han \cite{MENGXC09}).

A remarkable feature in figure \ref{xc} is that when $Z>0.02$, the
relation between the carbon abundance and the initial WD mass
significantly deviates from those of $Z\leq0.02$ and the deviation
increases with metallicity, i.e. for a given initial WD mass, a
high metallicity leads to a lower carbon abundance as shown in
figure \ref{structure}. This is similar to that found by Timmes et
al. (\cite{TIM03}), i.e. only when $Z>0.02$, the influence of
metallicity becomes significant. So, the discovery of Umeda et al.
(\cite{UMEDA99}) and Nomoto et al. (\cite{NOM99, NOM03}) is a
low-metallicity limit of the result found in this paper.

In the three-dimensional space of ($X_{\rm C}, M_{\rm core}, Z$),
the relations between the carbon abundance and the initial WD mass
with different metallicity are almost in a plane. We take the
$X_{\rm C}$ as a function of $M_{\rm core}+\alpha Z$ and use the
minimum $\chi^{\rm 2}$ method to find the best relation. We found
that a linear relation may well represent the relation when
$\alpha=3.5$. The best fitted relation is
\begin{equation}
 X_{\rm C}=0.5531-0.2528(M_{\rm core}+\alpha Z),
 \label{eqxc}
  \end{equation}
where $\chi^{\rm 2}_{\rm min}=5.16\times10^{\rm -2}$. We also try
to use a parabola to fit the relation between $X_{\rm C}$ and
$M_{\rm core}+\alpha Z$, but the improvement is slight, i.e.
$\chi^{\rm 2}_{\rm min}=4.84\times10^{\rm -2}$. Seen from the
equation \ref{eqxc}, both a high initial WD mass and a high
initial metallicity lead to a lower carbon abundance. In addition,
if the metallicity is low enough, its effect on the carbon
abundance can be neglected. But the effect of a high metallicity
is significant. For example, for a typical value of $M_{\rm i}=0.8
M_{\odot}$ (Meng, Chen \& Han \cite{MENGXC09}; Meng \& Yang
\cite{MENGYANG10a}), the uncertainty of $X_{\rm C}$ derived from
metallicity for $Z\leq0.02$ is less than 5\%, but may be as large
as 25\% for $Z>0.02$.

       \begin{figure}
   \centering
   \includegraphics[width=60mm,height=80mm,angle=270.0]{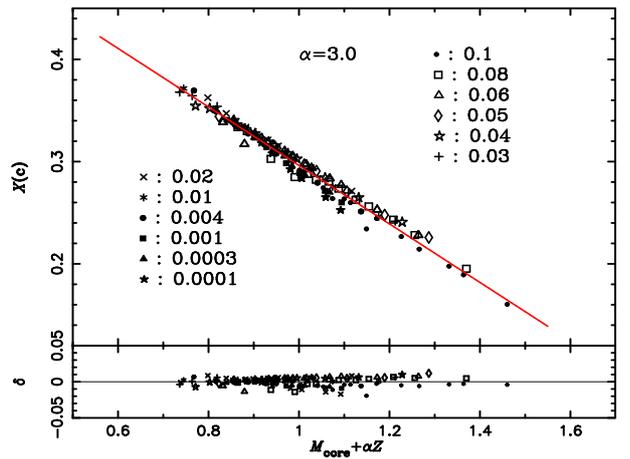}
   \caption{The relation between the total carbon abundance of
    WDs before SNe Ia explosion, the initial
WD mass and metallicity, where the points which may not contribute
to SNe Ia are cut off. The solid line is the best fitted linear
relation. The lower panel shows the difference between the points
and the fitted line.} \label{cut300}%
    \end{figure}

If the points which may not contribute to SNe Ia in Fig. \ref{xc}
are cut off, we even may obtain a more tight linear relation, i.e.
$X_{\rm C}=0.5824-0.2862(M_{\rm core}+\alpha Z)$ where
$\alpha=3.0$ and $\chi^{\rm 2}_{\rm min}=1.46\times10^{\rm -2}$
(see Fig. \ref{cut300}).

\section{Discussions and conclusions}\label{sect:4}
When one simulates the SN Ia explosion, the carbon abundance (the
C/O ratio) and metallicity are always taken as free parameters.
But neither of them has an ability to interpret the variation in
the mass of $^{\rm 56}$Ni ejected by SNe Ia, especially for
subluminous SNe Ia (such as 1991bg-like supernovae, Timmes et al.
\cite{TIM03}; R\"{o}pke et al. \cite{ROPKE06}; Bravo et al.
\cite{BRAVO10b}). However, many evidence shows the correlation
between the maximum luminosity of SNe Ia and metallicity. For
example, many groups noticed that subluminous SNe Ia occur
exclusive in massive galaxies (Neill et al. \cite{NEILL09};
Sullivan et al. \cite{SULLIVAN10}; Gonz\'{a}lez-Gait\'{a}n et al.
\cite{SONZALEZ10}; Lampeitl et al. \cite{LAMPEITL10}). Considering
the mass-metallicity relation of galaxies (Tremonti et al.
\cite{TREMONTI04}), subluminous SNe Ia favor more metal-rich
environments. In addition, some 1991T-like SNe Ia were discovered
in metal-poor environments\footnote{Even 2003fg-like SNe Ia which
are over-luminous events favor metal-poor environments
(Taubenberger et al. \cite{TAUBENBERGER10}), but the discussion
about these SNe Ia is beyond the scope of this paper since these
SNe Ia are over Chandrasekhar mass limit (Howell et al.
\cite{HOW06}) and this paper is based on the Chandrasekhar mass
model.} (Prieto et al. \cite{PRIETO08}; Badenes et al.
\cite{BADENES09}; Khan et al. \cite{KHAN10}). So, metallicity
should play a more significant role in a rather wide range than
that suggested in theory. The confliction between theory and
observations should be resolved.

In this paper, we found that when $Z>0.02$, the carbon abundance
before SN Ia explosion is affected by not only initial WD mass but
also metallicity. For a given initial WD mass, the higher the
metallicity, the lower the carbon abundance. The relation between
the carbon abundance, the initial WD mass and the metallicity may
be represented by a simple linear relation. Thus, when one
simulates the SN Ia explosion, the carbon abundance (the C/O
ratio) and metallicity would not be taken as free parameters.
Their effect on the amount of $^{\rm 56}$Ni is enhanced by each
other since the effect of a high metallicity and a low carbon
abundance on the amount of $^{\rm 56}$Ni is similar, i.e.
producing a lower amount of $^{\rm 56}$Ni (Nomoto et al.
\cite{NOM99, NOM03}; Timmes et al. \cite{TIM03}). So, our result
may account for the variation of the maximum luminosity of SNe Ia,
at least qualitatively providing a method to conquer the
confliction stated above. However, please keep in mind that
whether or not the combined action of the C/O ratio and
metallicity has enough ability to interpret the variation of the
maximum luminosity of SNe Ia still should be verified carefully.
\emph{So, we encourage someone to do a detailed numerical
simulation on this problem based on the result here}.

Furthermore, the central density at the moment of supernova
explosion may also play a role to some extent (H\"{o}flich et al.
\cite{HOFLIC10}; Krueger et al \cite{KRUEGER10}), and the carbon
abundance, metallicity and the central density all contribute to
the variation of the maximum luminosity of SNe Ia (R\"{o}pke et
al. \cite{ROPKE06b}). By a simple assumption that the carbon
abundance is the function of initial WD mass and the central
density is determined by initial WD mass and its cooling time,
Meng et al. (\cite{MENGYANGLI10}) noticed that the WDs with a high
carbon abundance usually have a lower central density at ignition,
while those having the highest central density at ignition
generally have a lower carbon abundance. Interestingly, the effect
of a high metallicity, a low C/O ratio and a high central density
on the amount of $^{\rm 56}$Ni is similar, i.e. producing a less
amount of $^{\rm 56}$Ni and thus a dimmer event (Nomoto et al.
\cite{NOM99,NOM03}; Timmes et al. \cite{TIM03}; Krueger et al
\cite{KRUEGER10}), then all of them could contribute the fact that
elliptical galaxies favor the dim SNe Ia (Hamuy \cite{HAM96}).
Although it is still unclear how the metallicity affects the
central density, we may hypothesize optimistically that it is
unreasonable that take the carbon abundance, metallicity and the
central density as free parameters when one simulates SN Ia
explosion, and these parameters may all contribute the production
of $^{\rm 56}$Ni (R\"{o}pke et al. \cite{ROPKE06b}), which could
be a key to open the origin of the Phillips relation. Moreover,
these parameters can be determined by the initial parameters of a
progenitor system, i.e. the WD mass, metallicity, orbital period
and secondary mass. For example, the C/O ratio is a function of
the initial WD mass and metallicity, while the central density of
a WD before supernova explosion is determined by accretion rate
and its cooling time before the onset of mass transfer, which is
related with the initial WD mass, secondary mass and period. Then
our study might provide a bridge linking the progenitor model and
explosion model of SNe Ia.

\begin{acknowledgements}
This work was partly supported by Natural Science Foundation of
China under grant no. 11003003 and the Project of the Fundamental
and Frontier Research of Henan Province under grant no.
102300410223, the Project of Science and Technology from the
Ministry of Education (211102) and the China Postdoctoral Science
Foundation funded project 20100480222.
\end{acknowledgements}

\newpage
%\appendix
\begin{appendix}
%\hspace{0.5mm}\\
\section[]{Some physical quantities for models}
In this paper, we calculated a large and fine grid models. Here,
we provide some physical quantities of the models such as CO core
mass, total carbon abundance X(C), initial main-sequence mass and
metallicity, which may be helpful for completing the resolution in
the field. Here, the boundary of the CO core was located at the
10\% of the maximum helium abundance in the helium shell (see the
left dotted line in Fig. \ref{structure}).

\begin{table*}[]
\caption[]{The CO core masses (in $M_{\odot}$) for different
metallicities (in $Z_{\odot}$, Row 1) and different initial masses
(Column 1, in $M_{\odot}$). The bars represent models whose final
fates are ONeMg white dwarfs.} \label{Tab:1}
\begin{center}
\begin{tabular}{ccccccccccccc}
\hline\noalign{\smallskip}%\scriptsize
       & 0.005  & 0.015  & 0.05   & 0.2    & 0.5    & 1.0    & 1.5    & 2.0    & 2.5    & 3.0    & 4.0    & 5.0  \\
\hline\noalign{\smallskip}
 1.00  & 0.6683 & 0.6490 & 0.6196 & 0.5786 & 0.5531 & 0.5366 & 0.5283 & 0.5272 & 0.5240 & 0.5292 & 0.5698 & 0.6358\\
 1.30  & 0.7033 & 0.6798 & 0.6433 & 0.5981 & 0.5698 & 0.5533 & 0.5438 & 0.5406 & 0.5403 & 0.5435 & 0.5844 & 0.6438\\
 1.50  & 0.7266 & 0.7007 & 0.6602 & 0.6112 & 0.5812 & 0.5634 & 0.5538 & 0.5503 & 0.5477 & 0.5609 & 0.5877 & 0.6450\\
 1.80  & 0.7562 & 0.7281 & 0.6846 & 0.6308 & 0.5985 & 0.5784 & 0.5687 & 0.5621 & 0.5659 & 0.5687 & 0.6036 & 0.6481\\
 2.00  & 0.7746 & 0.7488 & 0.7038 & 0.6437 & 0.6094 & 0.5881 & 0.5788 & 0.5749 & 0.5724 & 0.5774 & 0.6242 & 0.6698\\
 2.29  & 0.7725 & 0.7863 & 0.7454 & 0.6680 & 0.6284 & 0.6051 & 0.5953 & 0.5938 & 0.5943 & 0.6060 & 0.6698 & 0.6736\\
 2.51  & 0.8258 & 0.7898 & 0.7637 & 0.6939 & 0.6449 & 0.6196 & 0.6109 & 0.6093 & 0.6223 & 0.6378 & 0.6764 & 0.6973\\
 2.79  & 0.8319 & 0.8208 & 0.7961 & 0.7515 & 0.6761 & 0.6460 & 0.6373 & 0.6410 & 0.6648 & 0.6747 & 0.6856 & 0.7737\\
 2.99  & 0.8522 & 0.8420 & 0.8178 & 0.7640 & 0.7093 & 0.6718 & 0.6657 & 0.6729 & 0.6872 & 0.6894 & 0.7027 & 0.7777\\
 3.31  & 0.8884 & 0.8804 & 0.8570 & 0.7971 & 0.7921 & 0.7323 & 0.7120 & 0.7120 & 0.7115 & 0.7136 & 0.7377 & 0.7913\\
 3.51  & 0.9112 & 0.9059 & 0.8837 & 0.8187 & 0.7674 & 0.7411 & 0.7308 & 0.7292 & 0.7276 & 0.7289 & 0.7573 & 0.8208\\
 3.80  & 0.9500 & 0.9461 & 0.9258 & 0.8563 & 0.7978 & 0.7647 & 0.7550 & 0.7512 & 0.7526 & 0.7564 & 0.7899 & 0.8765\\
 3.98  & 0.9748 & 0.9740 & 0.9543 & 0.8801 & 0.8176 & 0.7805 & 0.7699 & 0.7652 & 0.7711 & 0.7761 & 0.8145 & 0.9095\\
 4.32  & 1.0252 & 1.0274 & 1.0159 & 0.9313 & 0.8620 & 0.8134 & 0.7983 & 0.7977 & 0.7992 & 0.8124 & 0.8619 & 0.9869\\
 4.52  & 1.0561 & --     & 1.0669 & 0.9661 & 0.8898 & 0.8367 & 0.8166 & 0.8167 & 0.8199 & 0.8351 & 0.8971 & --\\
 5.01  & --     & --     & --     & 1.0816 & 0.9733 & 0.9013 & 0.8725 & 0.8727 & 0.8833 & 0.9049 & 0.9983 & --\\
 5.50  & --     & --     & --     & --     & 1.0913 & 0.9869 & 0.9391 & 0.9431 & 0.9582 & 0.9956 & --     & --\\
 6.03  & --     & --     & --     & --     & --     & --     & 1.0448 & 1.0476 & 1.0710 & --     & --     & --\\
\noalign{\smallskip}\hline
\end{tabular}\end{center}
\end{table*}

\begin{table*}[]
\caption[]{The total carbon mass fraction of WDs before SNe Ia
explosion for different metallicities (in $Z_{\odot}$, Row 1) and
different initial masses (Column 1, in $M_{\odot}$). The bars
represent models whose final fates are ONeMg white dwarfs.}
\label{Tab:2}
\begin{center}
\begin{tabular}{ccccccccccccc}
\hline\noalign{\smallskip}%\scriptsize
       & 0.005  & 0.015  & 0.05   & 0.2    & 0.5    & 1.0    & 1.5    & 2.0    & 2.5    & 3.0    & 4.0    & 5.0  \\
\hline\noalign{\smallskip}
 1.00  & 0.3971 & 0.3964 & 0.3901 & 0.3886 & 0.3828 & 0.3775 & 0.3735 & 0.3560 & 0.3589 & 0.3543 & 0.3298 & 0.3048\\
 1.30  & 0.3949 & 0.3943 & 0.3898 & 0.3883 & 0.3825 & 0.3771 & 0.3726 & 0.3566 & 0.3578 & 0.3516 & 0.3294 & 0.2917\\
 1.50  & 0.3930 & 0.3923 & 0.3906 & 0.3881 & 0.3823 & 0.3770 & 0.3721 & 0.3563 & 0.3561 & 0.3364 & 0.3357 & 0.2917\\
 1.80  & 0.3879 & 0.3861 & 0.3878 & 0.3876 & 0.3821 & 0.3764 & 0.3714 & 0.3607 & 0.3480 & 0.3489 & 0.3337 & 0.2743\\
 2.00  & 0.3835 & 0.3820 & 0.3829 & 0.3871 & 0.3820 & 0.3761 & 0.3711 & 0.3550 & 0.3540 & 0.3474 & 0.3298 & 0.2638\\
 2.29  & 0.3653 & 0.3696 & 0.3741 & 0.3845 & 0.3821 & 0.3760 & 0.3704 & 0.3464 & 0.3501 & 0.3435 & 0.3028 & 0.2635\\
 2.51  & 0.3554 & 0.3520 & 0.3582 & 0.3795 & 0.3808 & 0.3757 & 0.3702 & 0.3547 & 0.3478 & 0.3386 & 0.2848 & 0.2598\\
 2.79  & 0.3417 & 0.3395 & 0.3408 & 0.3695 & 0.3765 & 0.3735 & 0.3680 & 0.3544 & 0.3443 & 0.3172 & 0.2880 & 0.2342\\
 2.99  & 0.3335 & 0.3327 & 0.3350 & 0.3493 & 0.3716 & 0.3702 & 0.3645 & 0.3520 & 0.3395 & 0.3243 & 0.2823 & 0.2444\\
 3.31  & 0.3184 & 0.3207 & 0.3235 & 0.3334 & 0.3512 & 0.3625 & 0.3541 & 0.3403 & 0.3221 & 0.3060 & 0.2804 & 0.2267\\
 3.51  & 0.3080 & 0.3110 & 0.3140 & 0.3275 & 0.3393 & 0.3472 & 0.3378 & 0.3274 & 0.3150 & 0.3040 & 0.2735 & 0.2144\\
 3.80  & 0.2925 & 0.2958 & 0.2984 & 0.3165 & 0.3303 & 0.3337 & 0.3262 & 0.3193 & 0.3094 & 0.2978 & 0.2625 & 0.1977\\
 3.98  & 0.2836 & 0.2870 & 0.2886 & 0.3078 & 0.3250 & 0.3294 & 0.3231 & 0.3156 & 0.3045 & 0.2939 & 0.2562 & 0.1893\\
 4.32  & 0.2651 & 0.2700 & 0.2714 & 0.2899 & 0.3111 & 0.3212 & 0.3151 & 0.3078 & 0.2964 & 0.2836 & 0.2432 & 0.1605\\
 4.52  & 0.2526 & --     & 0.2601 & 0.2794 & 0.3009 & 0.3153 & 0.3105 & 0.3026 & 0.2907 & 0.2770 & 0.2282 & --\\
 5.01  & --     & --     & --     & 0.2512 & 0.2753 & 0.2942 & 0.2928 & 0.2842 & 0.2706 & 0.2530 & 0.1952 & --\\
 5.50  & --     & --     & --     & --     & 0.2460 & 0.2712 & 0.2693 & 0.2647 & 0.2477 & 0.2281 & --     & --\\
 6.03  & --     & --     & --     & --     & --     & --     & 0.2416 & 0.2407 & 0.2258 & --     & --     & --\\
\noalign{\smallskip}\hline
\end{tabular}\end{center}
\end{table*}

\end{appendix}
\end{document}